\documentclass[aps,nofootinbib,preprintnumbers,twocolumn,superscriptaddress,prl,10pt]{revtex4-1}

\usepackage{amsmath}
\usepackage{amssymb}
\usepackage{slashed}
\usepackage{epsfig}
\usepackage{bbm,bm}
\usepackage{hyperref}
\usepackage{color}
\usepackage{comment}
\usepackage{amsfonts}
\usepackage{dsfont}

\def\bea{\begin{eqnarray}} 
\def\eea{\end{eqnarray}}
\def\be{\begin{equation}} 
\def\ee{\end{equation}} 
\def\ba{\begin{array}}
\def\ea{\end{array}}

\def\be{\begin{equation}}
\def\ee{\end{equation}}
\def\bea{\begin{eqnarray}}
\def\eea{\end{eqnarray}}

\def\Tr{\mbox{Tr}}

\begin{document}

\title{
Fractal Geometry of Higher Derivative Gravity
}

\author{Maximilian Becker}
\email{bemaximi@uni-mainz.de}
\affiliation{
Institute of Physics, Johannes Gutenberg University Mainz, Staudingerweg 7, D-55099 Mainz, Germany
}

\author{Carlo Pagani}
\email{carlo.pagani@lpmmc.cnrs.fr}
\affiliation{
Universit\'e Grenoble Alpes, CNRS, LPMMC, 38000 Grenoble, France
}
\affiliation{
Institute of Physics, Johannes Gutenberg University Mainz, Staudingerweg 7, D-55099 Mainz, Germany
}

\author{Omar Zanusso}
\email{omar.zanusso@unipi.it}
\affiliation{
Universit\`a di Pisa and INFN - Sezione di Pisa, Largo Bruno Pontecorvo 3, I-56127 Pisa, Italy
}

\begin{abstract}
We determine the scaling properties of geometric operators such as lengths, areas,
and volumes in models of higher derivative quantum gravity
by renormalizing appropriate composite operators.
We use these results to deduce the fractal dimensions of such
hypersurfaces embedded in a quantum spacetime at very small distances.
\end{abstract}

\pacs{}
\maketitle


It was shown a long time ago by Stelle \cite{Stelle:1976gc} that the action 
\begin{eqnarray}\label{eq:action4d-original}
 S[g] &= \int{\rm d}^4x \sqrt{-g}\left\{
  \frac{1}{f_2^2}\left(\frac{1}{3}R^2-R_{\mu\nu}R^{\mu\nu}\right)+\frac{1}{6f_0^2}R^2
 \right\}
\end{eqnarray}
is perturbatively renormalizable in four dimensions.
Stelle's model became even more attractive once it was shown to be asymptotically free
in the coupling $\lambda\equiv (f_2)^2$ \cite{Fradkin:1981iu,Fradkin:1981hx}.
However it was soon realized that the model is non-unitary because of the higher derivative propagator.
Nevertheless, solutions to this problem
were proposed early on \cite{Tomboulis:1977jk,Julve:1978xn,Salam:1978fd,Avramidi:1985ki} and invoked
a variety of ideas including, in particular, self-stabilization \cite{Salam:1978fd}
and the Lee-Wick mechanism \cite{Lee:1969fy}.
The interest toward higher derivative quantum gravity has resurged over the years
\cite{deBerredoPeixoto:2004if,deBerredoPeixoto:2003pj},
and recently has returned thanks to the appearance of new proposals
which are spiritual successors of the aforementioned ideas:
\emph{a}gravity
\cite{Salvio:2017qkx,Salvio:2018kwh} and a perturbatively unitary mechanism based on quantizing some degrees of freedom as \emph{fakeons} \cite{Anselmi:2018tmf,Anselmi:2018ibi}. 
Furthermore, it has been conjectured that unitarity is restored when the theory is assisted by a strongly coupled Yang-Mills theory \cite{Donoghue:2018izj} or when its coupling coefficient is larger than that of an added Einstein-Hilbert term \cite{Holdom:2016xfn}.

Alongside the development of higher derivative gravity the idea of non-perturbative renormalizability
of standard Einstein gravity has gained momentum and culminated in the asymptotic safety conjecture \cite{Weinberg:1980,Smolin:1981rm},
which has evidence based on non-perturbative renormalization group methods \cite{Reuter:1996cp,Souma:1999at}.
The status of the relation between asymptotically free Stelle's gravity
and asymptotically safe Einstein's gravity
has been debated by theorists for some time \cite{Codello:2006in,Niedermaier:2009zz}, especially
because the latter is believed to originate from the continuation of $(2+\varepsilon)$-gravity \cite{Christensen:1978sc}.
Explicit results based on mass-dependent regulators suggest that in four dimensions there could be two distinct universality classes \cite{Benedetti:2009rx,Groh:2011vn}.

The increasing attention toward Stelle's gravity and its high energy properties opens
the avenue to the discussion of its physical implications in the search for possible phenomenological signatures.
In fact, model specific implications have already been explored in various contexts \cite{Salvio:2019ewf,Anselmi:2019rxg,Anselmi:2018bra}.
The geometric characterization of the quantum theory of \eqref{eq:action4d-original},
which could be expected to have a fractal nature induced by radiative corrections, is, however, still lacking.
One straightforward tool to explore the geometry of quantum spacetimes
is the inclusion of composite
operators into the renormalization process which have a geometric meaning \cite{Pagani:2016dof,Becker:2018quq}
and, thus, can be used to deduce meaningful quantities such as,
for example, the fractal dimensions of embedded hypersurfaces of various (bare) dimensionalities.
This work is dedicated to the renormalization of some geometric operators
which allow one to read off such fractal dimensions.

Many quantum gravity scenarios predict that spacetime has a fractal
behavior at a very small scale, often implying that the dimension of spacetime is smaller than four.
Interestingly, this happens both in the asymptotic safety scenario
and in the causal dynamical triangulations approach (see \cite{Carlip:2017eud} for a comprehensive review).
It must be emphasized that there are, in principle, several possible working
definitions of the spacetime dimension. Examples include
the spectral dimension, the walk dimension and the Hausdorff dimension,
and all these definitions could give different estimates of the fractal dimension
\cite{Reuter:2011ah}.

\paragraph{Renormalization.}
We begin by recalling basic facts
on the renormalization of Stelle's gravity to set the stage for our results.
In the following, we adopt the notation of
\cite{Salvio:2017qkx},
which we
refer to for more details on the couplings' renormalization.
The bare action \eqref{eq:action4d-original} is the most general power-counting renormalizable action
constructed with curvature tensors of the metric $g_{\mu\nu}$ in four dimensions
modulo non-propagating boundary and topological terms.
It includes the square of the Weyl tensor $C^2=C_{\mu\nu\rho\theta}C^{\mu\nu\rho\theta}$
which is weighted by the coupling $f_2$ because it fulfills
$\int C^2=-2\int \left(\frac{1}{3}R^2-R_{\mu\nu}R^{\mu\nu}\right)$
by neglecting the contribution of the Euler characteristic.
The parametrization is chosen such that
the only term manifestly breaking the Weyl symmetry is $R^2$ which is weighted by the coupling $f_0$.
We refer to the conformally invariant limit $f_0\to\infty$ as Weyl's higher derivative gravity \cite{deBerredoPeixoto:2003pj}.
Operators with fewer derivatives, such as the scalar curvature $R$ which couples through Newton's contant or the
spacetime volume which couples through the cosmological constant, 
can, in principle, be included as relevant deformations of $S[g]$, but we will stick to (\ref{eq:action4d-original}).

To renormalize a path-integral constructed with the action \eqref{eq:action4d-original},
we adopt the background field method and split the metric in a background and a fluctuation part,
$ g_{\mu\nu} \to g_{\mu\nu} + h_{\mu\nu}$.
This split is used to fix the gauge, while the background metric is chosen to be flat, $g_{\mu\nu}=\eta_{\mu\nu}$, from now on,
which is enough to determine the counterterms.
We employ the following gauge-fixing action:
\begin{eqnarray}\label{eq:action4d-gf}
 S_{\mathrm{GF}}[h] &=& -\frac{1}{2\xi_g} \int{\rm d}^4x\, \chi_\mu\partial^2\chi^\mu
\end{eqnarray}
with $\chi_\mu=\partial^\nu\left(h_{\mu\nu}-c_g\frac{1}{2}\eta_{\mu\nu}{h_\alpha}^\alpha\right)$ and the two gauge-fixing parameters $\xi_g$ and $c_g$.

By adopting dimensional regularization and using minimal subtraction
one finds the beta functions
\begin{equation}\label{eq:beta-functions}
 \begin{split}
 \beta_{f_2}&=-\frac{1}{(4\pi)^2}\frac{133}{20}f_2^3 \,,\\
 \beta_{f_0}&=+\frac{1}{(4\pi)^2}\left(\frac{10}{12}\frac{f_2^4}{f_0^4}+\frac{5}{2}\frac{f_2^2}{f_0^2} +\frac{5}{12}\right)f_0^3 \, ,
 \end{split}
\end{equation}
where $\beta_{f_i}\equiv \frac{d}{d\log \mu}f_i$, with $\mu$ being the reference scale at which the renormalized couplings are defined.
Both beta functions admit Gaussian fixed points,
while the ratio $\omega =\frac{f_2^2}{2f_0^2}$ has a beta function $\beta_\omega$ with two non-trivial zeroes.
Thus, by setting $\omega$ to either fixed point one can obtain a perturbative series controlled solely
by $f_2$.
The conformal limit $f_0\to\infty$ is discontinuous because Weyl invariance
must be gauge fixed through the additional condition $h_\mu{}^\mu=0$ and the number of propagating degrees of freedom changes \cite{deBerredoPeixoto:2003pj}.
The renormalization group flow in this case becomes
\begin{eqnarray}\label{eq:beta-functions-conformal}
 \beta_{f_2} &=&-\frac{1}{(4\pi)^2}\frac{199}{30}f_2^3 \, .
\end{eqnarray}
Here, we explicitly assume that the beta function in the conformal limit is gauge independent as it is the case for the beta functions of Stelle's gravity \cite{avramidi}.

\paragraph{Scaling dimensions.}
We now introduce the scaling dimension of an embedded hypersurface and discuss its physical meaning.
The scaling dimension is sometimes used to guess the Hausdorff
dimension but may differ from it \cite{mandelbrot-book}. Let us consider the
volume of an $n$-dimensional surface $\sigma_{n}$ and denote it
by $V_{\sigma_{n}}$. Let us also assume that the volume is characterized
by some length $L$.\footnote{For instance, the volume is specified by an $n$-dimensional ball
of radius $L$ in the coordinate space.} Classically, we expect that $V_{\sigma_{n}}$ scales like $V_{\sigma_{n}}\sim L^{n}$.
In the quantum regime, however, the gravitational fluctuations might
change the classical scaling by modifying the scaling exponent via
an anomalous dimension $\gamma_n$, i.e.\ $V_{\sigma_{n}}\sim L^{n-\gamma_{n}}$.
In this case, we say that the $n$-dimensional surface
$\sigma_{n}$ has scaling dimension $n-\gamma_{n}$.
One can construct further definitions of scaling dimensions from the
building blocks $\sigma_{n}$. For instance, one may
measure the scaling dimension not in terms of the characteristic (coordinate) length
$L$, but rather in terms of the length of a given curve $\sigma_{1}$,
whose total length we denote by $V_{\sigma_{1}}$. Combining the scaling
behaviors $V_{\sigma_{n}}\sim L^{n-\gamma_{n}}$ and $V_{\sigma_{1}}\sim L^{1-\gamma_{1}}$,
one obtains that $V_{\sigma_{n}}\sim V_{\sigma_{1}}{}^{\frac{n-\gamma_{n}}{1-\gamma_{1}}}$,
which defines a new scaling exponent for the volume of $\sigma_{n}$.

\paragraph{Anomalous dimensions.}
Let us introduce the volume of $\sigma_{n}$ on the field theoretical side now. 
The induced metric on $\sigma_{n}$ is given by the pull-back of the spacetime
metric onto the surface, which we parametrize by $x^\mu(u)$ via the coordinates
$u^{a}$ with $a=1,\dots,n$. The pulled-back induced metric is given by
\begin{equation}
g_{ab}(u)=g_{\mu\nu}\left(x(u)\right)\frac{\partial x^{\mu}}{\partial u^{a}}\frac{\partial x^{\nu}}{\partial u^{b}}\,,\label{eq:defgab}
\end{equation}
and the volume of the submanifold $\sigma_{n}$ then is written as
\begin{align}
V_{\sigma_{n}} & \equiv\int_{\sigma_{n}}\sqrt{x^{*}g}\,=\,\int_{D}d^{n}u\sqrt{\det g_{ab}(u)}\,.\label{eq:vol}
\end{align}
It is easy to see that equation \eqref{eq:vol} for the case $n=1$ reproduces the
length of a given curve $x^{\mu}(u)$,
\begin{eqnarray*}
V_{\sigma_{1}} & = & \int du\,\sqrt{g_{\mu\nu}\left(x(u)\right)\dot{x}^{\mu}(u)\dot{x}^{\nu}(u)}\,.
\end{eqnarray*}
The induced volume element $\sqrt{g_{\sigma_{n}}}\equiv \sqrt{\det g_{ab}}$
is not present in the bare action \eqref{eq:action4d-original},
but it can be renormalized as a composite operator.
To do so we extend the action by adding a source $\zeta $ conjugate to the determinant of the pulled-back metric:
\begin{eqnarray}
S_{\cal O}[g,\zeta] \equiv \int d^n u\, \zeta \left(u\right) \sqrt{\det g_{ab} \left(u\right)}\,,
\end{eqnarray}
in which $S_{\cal O}[g,\zeta]$ is invariant under hypersurface-preserving diffeomorphism if the source $\zeta$ is transformed accordingly.

We denote the anomalous dimension of the composite operator by $\gamma_{\sigma_{n}}$
which at one loop is linear in $f_2^2$ and $f_0^2$.
The Callan-Symanzik equation for $\langle \sqrt{g_{\sigma_{n}}(u)}\rangle$
reads
\begin{eqnarray}
\left(\mu\partial_{\mu}+\beta_{f_2}\frac{\partial}{\partial f_2}+\beta_{f_0}\frac{\partial}{\partial f_0}
+\gamma_{\sigma_{n}}\right)\langle \sqrt{g_{\sigma_{n}}(u)} \rangle & = & 0\,.\label{eq:CS-eq-for-g_sigma_n-complete}
\end{eqnarray}
In the deep ultraviolet, i.e.\ for $f_2,f_0\rightarrow0$,
we can neglect the beta functions, which are cubic in the couplings,
and approximate equation \eqref{eq:CS-eq-for-g_sigma_n-complete}
as\footnote{We assume here that the expectation value of the composite operator
can be expanded perturbatively 
and that it is non-zero even at zero coupling, which is to be expected since gravity is naturally in the 
broken phase, i.e.~$\langle g_{\mu\nu} \rangle \approx \eta_{\mu\nu}\neq 0$.}
\begin{eqnarray}
\left(\mu\partial_{\mu}+\gamma_{\sigma_{n}}\right)\langle \sqrt{g_{\sigma_{n}}(u)} \rangle
& \approx & 0\,.\label{eq:CS-eq-for-g_sigma_n-approx}
\end{eqnarray}
Equation (\ref{eq:CS-eq-for-g_sigma_n-approx}) facilitates 
the scaling analysis of correlation functions in conjunction with 
dimensional analysis, which provides a further, independent equation.
In particular, since the metric is dimensionless,
one has
\begin{eqnarray}
\left(\mu\partial_{\mu}-u\partial_{u}\right)\langle \sqrt{g_{\sigma_{n}}(u)} \rangle & = & 0\,,\label{eq:dim-analysis}
\end{eqnarray}
in which we assume the
energy scale of interest is much bigger than all other
dimensionful quantities other than the coordinates (e.g.~any infrared mass).\footnote{Equivalently,
we assume that the correlation function depends only on 
the product $\left(u\mu \right)$.}
Combining equations (\ref{eq:CS-eq-for-g_sigma_n-approx}) and (\ref{eq:dim-analysis})
together, one obtains
\begin{eqnarray*}
\langle \sqrt{g_{\sigma_{n}}(u)} \rangle & \sim & u^{-\gamma_{\sigma_{n}}}\,.
\end{eqnarray*}
Thus, one can estimate the scaling behavior of $V_{\sigma_{n}}$
via
\begin{eqnarray*}
\langle V_{\sigma_{n}}\rangle\,=\,\int_{D}d^{n}u\,\langle \sqrt{g_{\sigma_{n}}(u)} \rangle & \sim & L^{n-\gamma_{\sigma_{n}}}\,,
\end{eqnarray*}
where $L$ is the characteristic length of the domain of integration.
This proves that at very high energies, or alternatively at very small scales,
the exponents coincide, $\gamma_n=\gamma_{\sigma_n}$,
and we can determine the scaling properties by field theoretic methods.

In the case of Stelle's gravity, the ultraviolet fixed-point is Gaussian; thus in the infinite energy
limit the anomalous dimensions observables are zero or, more precisely, radiative corrections to the scaling behavior are only logarithmic.
However, in a regime in which the coupling is sufficiently small, 
we are allowed to neglect the beta functions in \eqref{eq:CS-eq-for-g_sigma_n-complete}
and we encounter approximate scale invariance in which
a fractal-like behavior of geometrical volumes is present.\footnote{Notice
that, in the case $d=4-\varepsilon$, a non-trivial scaling behavior is present
even at the ultraviolet non-Gaussian fixed point.}
Let us note that, even if scaling dimensions are not generally universal away from a fixed point, in the present approximation both the anomalous
dimension and the couplings are independent of the renormalization scheme, implying that the approximate scale invariance is a 
physical effect.
This is the scaling regime located above any physical mass (including the Planck mass)
and inside the possibly infinite energy range of validity of \eqref{eq:action4d-original}, which displays scale-invariant (fractal) properties.

\paragraph{Geometric composite operators.} To explicitly derive the scaling exponent $\gamma_{\sigma_n}$,
we couple the composite operator of interest to a local source $\zeta(u)$
in \eqref{eq:action4d-original}.
At one loop a new divergence associated to the composite operator
can be computed by employing the standard trace-log formula for the
source-dependent effective action
\begin{eqnarray*}
\Gamma\left[g,\zeta\right] & = & S\left[g,\zeta\right]+\frac{1}{2}\Tr \log\left(\delta^{2}S\left[g,\zeta\right]\right)\,.
\end{eqnarray*}
The new divergences can be renormalized multiplicatively by
introducing a suitable counterterm 
for $S_{\cal O}[g,\zeta]$: $\int d^{n}u\,\zeta(u)Z_{\sigma_{n}}\sqrt{ g_{\sigma_n}(u)}$.
The anomalous dimension is then given by the coefficient of the pole
of $Z_{\sigma_{n}}$, which is computed by evaluating the one-point
function
\begin{align}
 \frac{\delta\Gamma}{\delta\zeta}\Bigr|_{\zeta=0}  = 
 &Z_{\sigma_{n}}\sqrt{ g_{\sigma_n}}+\frac{1}{2}\Tr\left[{\cal G}\cdot\left(\delta^{2}\sqrt{ g_{\sigma_n}}\right)\right]\,,\label{eq:1-loop-formula-expanded}
\end{align}
in which we keep all terms up to ${O}(f_2^2)$ and ${O}(f_0^2)$.
Here, ${\cal G}$ denotes the gauge-fixed graviton propagator
%
\begin{equation*} 
\begin{split}
{\cal G}^{\rho\sigma}_{\mu\nu}=\frac{\mathrm{i}}{k^4}\Big\{&-2 f_2^2 P^{(2)}+f_0^2\Bigl[P^{(0)}
+\frac{\sqrt{3}\,c_g T^{(0)}}{2-c_g}
\\& +\frac{3c_g^3 P^{(0\omega)}}{(2-c_g)^2} \Bigr]
+2\xi_g\Bigl[P^{(1)}+\frac{2 P^{(0\omega)}}{(2-c_g)^2} \Bigr] \Bigr\}^{\rho\sigma}_{\mu\nu} \, ,
\end{split}
\end{equation*}
in which $P^{(0)}$, $P^{(1)}$, $P^{(2)}$, $P^{(0\omega)}$ and $T^{(0)}$ are spin-projectors whose exact form is stated in \cite{Salvio:2017qkx}.

The explicit computation of \eqref{eq:1-loop-formula-expanded} for Stelle's model gives
\begin{equation} \label{eq:gamma-determination}
\begin{split}
\gamma_{\sigma_n}=&\frac{1}{(4\pi)^2}
\frac{n}{288}\Bigl\{
20(2-5n) f_2^2
-\Bigl[
(11n-26)
\\&
+\frac{6c_g(7n-10)}{2-c_g}
-\frac{9c_g^2(n+2)}{(2-c_g)^2}
\Bigr]f_0^2 
\\
&-12\Bigl[
(2-5n)-\frac{(n+2)}{(2-c_g)^2}
\Bigr] \xi_g \Bigr\}\, ,
\end{split}
\end{equation}
which is linear in both couplings $f_0^2$ and $f_2^2$ and in the gauge-fixing parameter $\xi_g$.
The anomalous dimension (\ref{eq:gamma-determination}) is scheme independent at this order.\footnote{A
systematic improvement of our estimate may be based on either pushing
perturbation theory to higher order or on applying the functional renormalization group together with optimization techniques; see \cite{Litim:2001up,Balog:2019rrg} for their application in statistical mechanics.}
The physical interpretation of this result is that all modes propagating in the gauge-fixed
propagator contribute to \eqref{eq:gamma-determination}, including both, gauge-invariant spin-$2$ (graviton) and scalar
modes as well as the unphysical vector and pseudoscalar ones.

The gauge dependence of \eqref{eq:gamma-determination} is to be expected because
embedded hypersurfaces break diffeomorphism invariance and thus, strictly speaking,
are not true observables. 
%
We could circumvent this problem by constructing a gauge-invariant observable 
which combines the volume of an hypersurface with an observable amplitude
such that the various gauge dependencies cancel each other.
This typically results in very non-local observables such as the correlation length at fixed geodesic length.
A similar program works nicely in $2d$ quantum gravity \cite{KPZ88,DDK88} where computations are typically performed
in the conformal gauge (to the best of our knowledge there is no study exploring the explicit gauge dependence cancellation).
However, the problem of constructing interesting and meaningful gauge-invariant observables in four dimensional
quantum gravity is a long-standing one 
\cite{DeWitt:1962cg,Rovelli:1990ph,Giddings:2005id,Ambjorn:1997di,Hamber:2009zz,Rovelli:2004tv},
and is beyond the scope of this work.

In order to find a simpler workaround we first notice (cf.\ \cite{Salvio:2018kwh}) that, in the \emph{physical} gauge $c_g=\xi_g=0$, only the gauge-invariant, hence physical, modes propagate.
The physical gauge is often associated to the unique Vilkovisky-de Witt effective action
in which a nontrivial connection in field space ensures
that only physical modes are integrated over and
that the effective action is gauge independent \cite{Parker:2009uva}. 
Importantly, this gauge also ensures the vanishing of \eqref{eq:gamma-determination} at the Gaussian fixed-point.
Therefore, from now on, we work in the gauge $c_g=\xi_g=0$ 
and assume that the physical scaling of hypersurfaces is
given by this gauge, in which only gauge-invariant degrees of freedom
propagate.
(See also \cite{Wetterich:2017aoy,Wetterich:2019zdo} for a related approach where only physical degrees
of freedom are used.)

\paragraph{Anomalous scaling in Stelle's gravity.}
On the basis of the fact that only physical modes propagate,
we argue that the physical gauge limit of \eqref{eq:gamma-determination} gives a reliable estimate of the scaling dimension
of an embedded hypersurface.
In the limit $\xi_g = c_g= 0$ the anomalous dimension in terms of the couplings $f_2$ and $\omega=\frac{f_2^2}{2f_0^2}$ reads
\begin{equation}\label{eq:anomalous-dimension-stelle}
\begin{split}
\gamma_{\sigma_n}&=\frac{n}{(4\pi)^2}
\frac{f_2^2}{576}\Bigl\{
40(2-5n)+\frac{1}{\omega}
(26-11n)
\Bigr\}\,,
\end{split}
\end{equation}
which can be used in \eqref{eq:CS-eq-for-g_sigma_n-complete} and constitutes one of the main results of this letter.

In four dimensions, the value of $\gamma_{\sigma_n}$ depends on the
scale of the renormalization group flow at which the couplings are located.
The anomalous dimension vanishes at the fixed point but is non-zero as soon as we move away from it. Note that this feature will hold in the physical gauge for any theory exhibiting a
fixed point which is Gaussian.
For sufficiently small values of the couplings, we find approximate scale invariance characterized
by an effective fractal dimension of the geometric operators.
More precisely, if $f_2^2>0$ it is straightforward to check that the anomalous dimension is
positive or negative depending on the value of $n$. 
For $n=1,2$, i.e.~lengths and areas,
the anomalous dimension is positive for $\frac{1}{\omega}>\frac{80 - 200 n}{-26 + 11 n}$,
while for $n=3,4$, i.e.~three- and four-volumes, for
$\frac{1}{\omega}<\frac{80 - 200 n}{-26 + 11 n}$.
It follows that quantum fluctuations affect hypersurfaces of different dimensions
in different manners: A length can effectively decrease its scaling dimension, while
the opposite happens for a three-volume.
Compared to other models of quantum gravity, which often display only dimensional reduction,
this behavior is very peculiar to higher derivative quantum gravity.

As already mentioned, in $d=4-\varepsilon$ the theory exhibits two nontrivial ultraviolet fixed points
which are solutions of $-\varepsilon f_2+2\beta_{f_2}=0$ and $\beta_\omega=0$.
The scaling of geometric operators at
such fixed points is characterized by $\gamma_{\sigma_n}$.
The system has two non-Gaussian solutions because the equation $\beta_\omega=0$ 
has two roots $\omega_{*,1}=-0{.}0229$ and $\omega_{*,2}=-5{.}4671$.
We label them by
\begin{eqnarray}\label{eq:fixed-points-stelle}
(f_2^2,\omega)_{*,1} &=&(-11{.}8732\,\varepsilon,-0{.}0229) \,, \\
(f_2^2,\omega)_{*,2} &=&(-11{.}8732\,\varepsilon,-5{.}4671) \, .
\end{eqnarray}
We argue that the more important solution is the first one, because it is fully ultraviolet attractive and
because the second one was shown to lead to a non-positive ghost inverse propagator \cite{Groh:2011vn}.
The values of $\gamma_{\sigma_n}$ at both fixed points can be exactly calculated, but for compactedness we give their numerical approximations in Tab.~\ref{table:fpgammaw}.

\begin{table}[b]
\caption{Leading estimates of the anomalous dimension \eqref{eq:anomalous-dimension-stelle} at the fixed points
 in $d=4-\varepsilon$ spacetime dimensions which correct the scaling of all hypersurfaces of dimension $n$ lower than four.\\}
\centering
\renewcommand{\arraystretch}{1.5}
\begin{tabular}{l r r r r}
\hline\hline 
 & $n=1$ & $n=2$ & $n=3$ & $n=4$ \\
 $\left.\gamma_{\sigma_n}\right|_{*,1}$ & $ \phantom{+}0{.}1012\varepsilon$ & $ \phantom{+}0{.}1291\varepsilon$ & $ \phantom{+}0{.}0839\varepsilon$ & $-0{.}0348\varepsilon$ \\
 $\left.\gamma_{\sigma_n}\right|_{*,2}$ & $ \phantom{+}0{.}0160\varepsilon$ & $ \phantom{+}0{.}0837\varepsilon$ & $ \phantom{+}0{.}2031\varepsilon$ & $ \phantom{+}0{.}3742\varepsilon$ \\
 $\left.\gamma_{\sigma_n}\right|_{*,w}$ & $ \phantom{+}0{.}0157\varepsilon$ & $ \phantom{+}0{.}0838\varepsilon$ & $ \phantom{+}0{.}2041\varepsilon$ & $ \phantom{+}0{.}3769\varepsilon$\\
[1ex] 
\hline
\end{tabular}
\label{table:fpgammaw}
\end{table}

\paragraph{Anomalous scaling in Weyl-squared gravity.}
The computation of $\gamma_{\sigma_{n}}$ in the Weyl-invariant case
goes along the same lines as in the case of Stelle's theory with the difference
being that the propagator in \eqref{eq:1-loop-formula-expanded} contains only spin-$2$ propagating modes. Unfortunately at two loops the conformal symmetry is anomalous \cite{deBerredoPeixoto:2003pj} and, consequently, radiative corrections will also generate the propagating scalar mode;
nevertheless Weyl theory is still relevant for situations in which the conformal symmetry is
approximately realized \cite{Rachwal:2018gwu}.
At one loop, the propagator is given by $\mathcal G$ with $f_0^2=\xi_g=0$ which also employs the additional gauge-fixing condition ${h_\mu}^\mu=0$.
The explicit result for the anomalous dimension then reads
\begin{equation}\label{eq:anomalous-dimension-weyl}
\begin{split}
\gamma_{\sigma_n}&= 
-\frac{1}{(4\pi)^{2}}\frac{10}{144}(5n^{2}-2n)\,f_2^2 \,.
\end{split}
\end{equation}
In $d=4$ the anomalous dimension is zero at the fixed point but non-zero in its neighbourhood. 
It is straightforward to see that the sign of the correction depends on the sign of $f_2^2$
for hypersurfaces of dimension $1<n\leq 4$ and not on $n$ itself.
In $d=4-\varepsilon$ there is only one solution of $-\varepsilon f_2 +2\beta_{f_2}=0$, with the beta function \eqref{eq:beta-functions-conformal}, in this case
and we include the estimates for the scaling dimensions in the last line of Tab.~\ref{table:fpgammaw} with the label $w$ for Weyl. 
The anomalous dimension at the non-trivial fixed point in $d=4-\varepsilon$
implies an effective dimensional reduction in the UV
for $\varepsilon>0$.

\paragraph{Summary and future prospects.}
Since by nature gravity is a geometrical theory, we believe that it is natural to investigate
the quantum properties of geometrical objects, such as lines, areas, and volumes
in quantum gravity.
In this letter we have considered the quantum properties of such geometric operators
in higher derivative gravity for the first time.
More precisely, we have computed the scaling properties of these geometric operators 
in Stelle's and Weyl theories in $d=4$ and $d=4-\varepsilon$.
 
For the most physically relevant case corresponding to $d=4$,
we have found that these geometric operators display a peculiar scaling behavior:
At the Gaussian fixed point the scaling is purely classical while moving away
from it we have a regime of approximate scale invariance in which the effective dimension is fractal.
The nature of this fractal behavior depends on the couplings and, thus, on the precise
scale at which the operators are observed.

Remarkably, similar geometric operators can be defined also in other approaches to quantum gravity,
such as loop quantum gravity, and causal dynamical triangulations \cite{Rovelli:2004tv,Ambjorn:2012jv}.
Therefore this work paves the way to a possible comparison among the predictions of
all different quantum gravity models, now including higher derivative quantum gravity.
In general, the comparison of our present results with \cite{Pagani:2016dof,Becker:2018quq} corroborates the intuition that Stelle's asymptotically free gravity and Einstein's asymptotically safe gravity are two distinct universality classes which are characterized by rather different fractal behaviors of the respective geometrical operators.

The biggest open issue of the approach presented here is to find a gauge-invariant generalization
of our results on the scaling dimensions.
As a matter of fact, this search overlaps with the quest for meaningful
gauge invariant observables in theories of quantum gravity.
Our approach offers a shortcut based on the choice of propagating only the physical, i.e., gauge invariant degrees of freedom
in full analogy with the Vilkovisky-de Witt formalism.
However, to what extent our approach is valid should be tested further.
In any case, let us emphasize that the approach developed here can also serve to study
fully fledged diffeomorphism-invariant observables. 
For instance, one could consider a correlation function at fixed geodesic length between two operators:
$\langle \int_x \int_y O(x)O(y) \delta \left(\ell_g-r\right) \rangle$ with $\ell_g$ being the geodesic length.
Performing a scaling analysis of such correlation functions involves the computation
of the scaling dimension of the geodesic length itself,
which can be computed in a way similar to the one outlined in this work.


\smallskip

\paragraph*{Acknowledgments.}

MB is supported by DFG Grant No.\ RE 793/8-1.
OZ was supported by the DFG Grant No.\ ZA 958/2-1
during the first stages of this work.



\begin{thebibliography}{99}







\bibitem{Stelle:1976gc} 
  K.~S.~Stelle,
  Phys.\ Rev.\ D {\bf 16}, 953 (1977).
  
\bibitem{Fradkin:1981iu} 
  E.~S.~Fradkin and A.~A.~Tseytlin,
  Nucl.\ Phys.\ B {\bf 201}, 469 (1982).
  
\bibitem{Fradkin:1981hx} 
  E.~S.~Fradkin and A.~A.~Tseytlin,
  Phys.\ Lett.\  {\bf 104B}, 377 (1981).
  
\bibitem{Julve:1978xn} 
  J.~Julve and M.~Tonin,
  Nuovo Cim.\ B {\bf 46}, 137 (1978).
  
\bibitem{Tomboulis:1977jk} 
  E.~Tomboulis,
  Phys.\ Lett.\  {\bf 70B}, 361 (1977).
  
\bibitem{Salam:1978fd} 
  A.~Salam and J.~A.~Strathdee,
  Phys.\ Rev.\ D {\bf 18}, 4480 (1978).

\bibitem{Avramidi:1985ki} 
  I.~G.~Avramidi and A.~O.~Barvinsky,
  Phys.\ Lett.\  {\bf 159B}, 269 (1985).
  
\bibitem{Lee:1969fy} 
  T.~D.~Lee and G.~C.~Wick,
  Nucl.\ Phys.\ B {\bf 9}, 209 (1969).
  
\bibitem{deBerredoPeixoto:2004if} 
  G.~de Berredo-Peixoto and I.~L.~Shapiro,
  Phys.\ Rev.\ D {\bf 71}, 064005 (2005)
  [hep-th/0412249].
  
\bibitem{deBerredoPeixoto:2003pj} 
  G.~de Berredo-Peixoto and I.~L.~Shapiro,
  Phys.\ Rev.\ D {\bf 70}, 044024 (2004)
  [hep-th/0307030].

\bibitem{Salvio:2017qkx} 
  A.~Salvio and A.~Strumia,
  Eur.\ Phys.\ J.\ C {\bf 78}, no. 2, 124 (2018)
  [arXiv:1705.03896 [hep-th]].
  
\bibitem{Salvio:2018kwh} 
  A.~Salvio, A.~Strumia and H.~Veerm{\"a}e,
  Eur.\ Phys.\ J.\ C {\bf 78}, no. 10, 842 (2018)
  [arXiv:1808.07883 [hep-th]].
  
\bibitem{Anselmi:2018tmf} 
  D.~Anselmi and M.~Piva,
  JHEP {\bf 1811}, 021 (2018)
  [arXiv:1806.03605 [hep-th]].
  
\bibitem{Anselmi:2018ibi} 
  D.~Anselmi and M.~Piva,
  JHEP {\bf 1805}, 027 (2018)
  [arXiv:1803.07777 [hep-th]].
  
  \bibitem{Donoghue:2018izj}
      J.~F.~Donoghue and G.~Menezes,
  Phys.\ Rev.\ D {\bf 97}, 126005 (2018)
    [arXiv:1804.04980 [hep-th]].

\bibitem{Holdom:2016xfn}
       B.~Holdom and J.~Ren,
        Int.\ J.\ Mod.\ Phys.\ {\bf D25}, 1643004 (2016)
    [arXiv:1605.05006 [hep-th]].
  
\bibitem{Weinberg:1980}  
  S.~Weinberg. 1980.
  in General Relativity, an Einstein Centenary Survey, S.~W.~Hawking and W.~Israel (eds.) Cambridge University Press, pg.~790
  
\bibitem{Smolin:1981rm} 
  L.~Smolin,
  Nucl.\ Phys.\ B {\bf 208}, 439 (1982).
  
\bibitem{Reuter:1996cp} 
  M.~Reuter,
  Phys.\ Rev.\ D {\bf 57}, 971 (1998)
  [hep-th/9605030].
  
\bibitem{Souma:1999at} 
  W.~Souma,
  Prog.\ Theor.\ Phys.\  {\bf 102}, 181 (1999)
  [hep-th/9907027].

\bibitem{Codello:2006in} 
  A.~Codello and R.~Percacci,
  Phys.\ Rev.\ Lett.\  {\bf 97}, 221301 (2006)
  [hep-th/0607128].
  
\bibitem{Niedermaier:2009zz} 
  M.~R.~Niedermaier,
  Phys.\ Rev.\ Lett.\  {\bf 103}, 101303 (2009).
  
\bibitem{Christensen:1978sc} 
  S.~M.~Christensen and M.~J.~Duff,
  Phys.\ Lett.\  {\bf 79B}, 213 (1978).
  
\bibitem{Benedetti:2009rx} 
  D.~Benedetti, P.~F.~Machado and F.~Saueressig,
  Mod.\ Phys.\ Lett.\ A {\bf 24}, 2233 (2009)
  [arXiv:0901.2984 [hep-th]].
  
\bibitem{Groh:2011vn} 
  K.~Groh, S.~Rechenberger, F.~Saueressig and O.~Zanusso,
  PoS EPS {\bf -HEP2011}, 124 (2011)
  [arXiv:1111.1743 [hep-th]].

\bibitem{Salvio:2019ewf} 
  A.~Salvio,
  Phys.\ Rev.\ D {\bf 99}, no. 10, 103507 (2019)
  [arXiv:1902.09557 [gr-qc]].
  
\bibitem{Anselmi:2019rxg} 
  D.~Anselmi,
  JHEP {\bf 1904}, 061 (2019)
  [arXiv:1901.09273 [gr-qc]].
  
\bibitem{Anselmi:2018bra} 
  D.~Anselmi,
  Class.\ Quant.\ Grav.\  {\bf 36}, 065010 (2019)
  [arXiv:1809.05037 [hep-th]].
  
\bibitem{Pagani:2016dof} 
  C.~Pagani and M.~Reuter,
  Phys.\ Rev.\ D {\bf 95}, no. 6, 066002 (2017)
  [arXiv:1611.06522 [gr-qc]].
  
\bibitem{Becker:2018quq} 
  M.~Becker and C.~Pagani,
  Phys.\ Rev.\ D {\bf 99}, no. 6, 066002 (2019)
  [arXiv:1810.11816 [gr-qc]].
  
\bibitem{Carlip:2017eud} 
  S.~Carlip,
  Class.\ Quant.\ Grav.\  {\bf 34}, no. 19, 193001 (2017)
  [arXiv:1705.05417 [gr-qc]].
  
\bibitem{Reuter:2011ah} 
  M.~Reuter and F.~Saueressig,
  JHEP {\bf 1112}, 012 (2011)
  [arXiv:1110.5224 [hep-th]].
  
\bibitem{avramidi} 
    I.G. Avramidi, Ph.D.\ thesis (1986) [hep-th/9510140]
  
\bibitem{mandelbrot-book}   
 B.~B.~Mandelbrot, ``The geometry of nature'', W.~H.~Freeman and Co.~(1982), USA.
 
\bibitem{Litim:2001up} 
  D.~F.~Litim,
  Phys.\ Rev.\ D {\bf 64}, 105007 (2001)
  doi:10.1103/PhysRevD.64.105007
  [hep-th/0103195].
  
\bibitem{Balog:2019rrg} 
  I.~Balog, H.~Chaté, B.~Delamotte, M.~Marohnic and N.~Wschebor,
  Phys.\ Rev.\ Lett.\  {\bf 123}, no. 24, 240604 (2019)
  [arXiv:1907.01829 [cond-mat.stat-mech]].

%
\bibitem{KPZ88}
  A.~M.~Polyakov,
  Mod.\ Phys.\ Lett.\ A {\bf 2}, 893 (1987);\\
  V.~G.~Knizhnik, A.~M.~Polyakov and A.~B.~Zamolodchikov,
  Mod.\ Phys.\ Lett.\ A {\bf 3}, 819 (1988).
  %

\bibitem{DDK88}
  F.~David,
  Mod.\ Phys.\ Lett.\ A {\bf 3}, 1651 (1988);\\
  J.~Distler and H.~Kawai,
  Nucl.\ Phys.\ B {\bf 321}, 509 (1989).
 %
  
\bibitem{DeWitt:1962cg} 
  B.~S.~DeWitt,
  ``The Quantization of geometry'',
  in ``{\it Gravitation: An introduction to current research}''.
  Wiley, 1962, chapter 8, pp.\ 266-381.
  
\bibitem{Rovelli:1990ph} 
  C.~Rovelli,
  Class.\ Quant.\ Grav.\  {\bf 8}, 297 (1991).
  
\bibitem{Giddings:2005id} 
  S.~B.~Giddings, D.~Marolf and J.~B.~Hartle,
  Phys.\ Rev.\ D {\bf 74}, 064018 (2006)
  [hep-th/0512200].
  
\bibitem{Ambjorn:1997di} 
  J.~Ambj{\o}rn, B.~Durhuus and T.~Jonsson,
  ``{\it Quantum Geometry : A Statistical Field Theory Approach}'',
  Cambridge University Press, Cambridge (UK), (1997)
  
\bibitem{Hamber:2009zz} 
  H.~W.~Hamber,
 ``{\it Quantum gravitation: The Feynman path integral approach}'',
  Springer,  Berlin Germany (2009).
  
\bibitem{Rovelli:2004tv} 
  C.~Rovelli,
  ``Quantum gravity'',
  Cambridge University Press, Cambridge (2004). 
  
\bibitem{Parker:2009uva} 
  L.~E.~Parker and D.~Toms,
  ``Quantum Field Theory in Curved Spacetime : Quantized Field and Gravity'',
  Cambridge University Press, (2009) UK.
  
\bibitem{Wetterich:2017aoy} 
  C.~Wetterich,
  Nucl.\ Phys.\ B {\bf 934}, 265 (2018)
  [arXiv:1710.02494 [hep-th]].
  
\bibitem{Wetterich:2019zdo} 
  C.~Wetterich and M.~Yamada,
  Phys.\ Rev.\ D {\bf 100}, no. 6, 066017 (2019)
  [arXiv:1906.01721 [hep-th]].
  
\bibitem{Rachwal:2018gwu} 
  L.~Rachwa\l,
  Universe {\bf 4}, no.\ 11, 125 (2018)
  [arXiv:1808.10457 [hep-th]].
  
\bibitem{Ambjorn:2012jv} 
  J.~Ambjorn, A.~Goerlich, J.~Jurkiewicz and R.~Loll,
  Phys.\ Rept.\  {\bf 519}, 127 (2012)
  [arXiv:1203.3591 [hep-th]].
 

  


  
\end{thebibliography}
\end{document}